\documentclass[english]{article}
\usepackage[T1]{fontenc}
\usepackage[latin9]{inputenc}
\usepackage{color}
\usepackage{float}
\usepackage{amsthm}
\usepackage{amsmath}
\usepackage{amssymb}
\usepackage{wasysym}
\usepackage{graphicx}

\makeatletter
\newcommand{\lyxaddress}[1]{
\par {\raggedright #1
\vspace{1.4em}
\noindent\par}
}
\theoremstyle{plain}
\newtheorem{thm}{\protect\theoremname}
  \theoremstyle{definition}
  \newtheorem{example}[thm]{\protect\examplename}
  \theoremstyle{plain}
  \newtheorem{lem}[thm]{\protect\lemmaname}
  \theoremstyle{definition}
  \newtheorem{defn}[thm]{\protect\definitionname}
\newenvironment{lyxlist}[1]
{\begin{list}{}
{\settowidth{\labelwidth}{#1}
 \setlength{\leftmargin}{\labelwidth}
 \addtolength{\leftmargin}{\labelsep}
 }}
{\end{list}}

\usepackage{graphicx}
\makeatletter
\providecommand{\bigsqcapa}{%
  \mathop{%
    \mathpalette\@updown\bigsqcup
  }%
}
\newcommand*{\@updown}[2]{%
  \rotatebox[origin=c]{180}{$\m@th#1#2$}%
}
\makeatother
\usepackage{url}
\usepackage{fullpage}
\usepackage{color}
\usepackage[colorlinks,naturalnames,citecolor=blue,urlcolor=blue]{hyperref}

\makeatother

\usepackage{babel}
  \providecommand{\definitionname}{Definition}
  \providecommand{\examplename}{Example}
  \providecommand{\lemmaname}{Lemma}
\providecommand{\theoremname}{Theorem}

\begin{document}

\title{Refinement Calculus of Reactive Systems}

\author{Viorel Preoteasa$^{1}$ and Stavros Tripakis$^{1,2}$}

\maketitle

\lyxaddress{$^{1}$Aalto University, Finland. \\
$^{2}$University of California, Berkeley, USA.}

\global\long\def\bool{\mathsf{Bool}}
\global\long\def\until{\;\mathsf{U}\;}
\global\long\def\always{\Box\,}
\global\long\def\event{\Diamond\,}
\global\long\def\rel{\mathsf{Rel}}
\global\long\def\skip{\mathsf{Skip}}
\global\long\def\nat{\mathsf{Nat}}

\global\long\def\mtran{\mathsf{MTran}}
\global\long\def\comp{\;;\,}
\global\long\def\assert#1{\{#1\}}
\global\long\def\pfail{\mathsf{fail}}
\global\long\def\magic{\mathsf{Magic}}
\global\long\def\fail{\mathsf{Fail}}

\global\long\def\assume#1{[#1]}
\global\long\def\pred{\mathsf{Pred}}
\global\long\def\type{\mathsf{Type}}
\global\long\def\demonic#1#2#3{[#1\leadsto#2 \;|\; #3]}

\global\long\def\nex{\fullmoon\,}
\global\long\def\onlynow{\mathsf{only\mbox{-}now}}
\global\long\def\angelic#1#2#3{\{#1\leadsto#2 \;|\;#3\}}

\global\long\def\fals{\mathsf{false}}
\global\long\def\tru{\mathsf{true}}
\global\long\def\grd{\mathsf{grd}}
\global\long\def\hide{\mathsf{hide}}
\global\long\def\inp{\mathsf{in}}
\global\long\def\gsystem#1{\{#1]}

\global\long\def\system#1#2{\{#1\ |\ #2]}

\global\long\def\proj{\mathsf{proj}}
\global\long\def\local{\mathsf{local}}
\global\long\def\counter{\mathsf{counter}}
\global\long\def\coun{\mathsf{count}}
\global\long\def\bcount{\mathsf{bcounter}}

\global\long\def\leads{\mathsf{\; L\;}}
\global\long\def\run{\mathsf{run}}
\global\long\def\localsys{\mathsf{LocalSys}}
\global\long\def\glocalsys{\mathsf{GrdLocalSys}}

\global\long\def\havoc{\mathsf{Havoc}}
\global\long\def\assertlive{\mathsf{AssertLive}}
\global\long\def\livehavoc{\mathsf{LiveHavoc}}
\global\long\def\reqresp{\mathsf{ReqResp}}

\global\long\def\gprec{\mathsf{gprec}}
\global\long\def\grel{\mathsf{grel}}
\global\long\def\ltran{\mathsf{localtran}}

\global\long\def\nxt#1{#1^{1}}
\global\long\def\lglsys#1{\{\!|#1]\!]}
\global\long\def\leadstost#1#2#3{#1\leadsto#2\leadsto#3}

\global\long\def\bigsqcap{\bigsqcapa}

\global\long\def\abullet{\mathbin{\hbox{\raise.1ex\hbox{\scriptsize\ensuremath{\bullet}}}}}

\global\long\def\bbullet{\mathbin{\hbox{\raise.22ex\hbox{\tiny\ensuremath{\bullet}}}}}

\global\long\def\sep{\mathbin{\hbox{\raise.22ex\hbox{\tiny\ensuremath{\bullet}}}}}
\global\long\def\bsys{\{\!|}
\global\long\def\esys{]\!]}

\global\long\def\lsysa#1#2#3#4{\bsys\,#1\; |\;#2\; |\;#3\; |\;#4\,\esys}

\global\long\def\lsysb#1#2#3{\bsys\,#1\; |\;#2 \; |\;#3\,\esys}

\global\long\def\lsysc#1#2{\bsys\,#1\; |\;#2\,\esys}

\global\long\def\lsysd#1{\bsys\,#1\,\esys}

\global\long\def\rcomp{\odot}

\global\long\def\ifs#1#2#3{\mathsf{if\ }#1\mathsf{\ then\ }#2\mathsf{\ else\ }#3}
\global\long\def\div{\mathsf{Div}}
\global\long\def\illegal{\mathsf{illegal}}

\global\long\def\relstcomp{\mathbin{\circ\circ}}

\begin{abstract}
Refinement calculus is a powerful and expressive tool for reasoning
about sequential programs in a compositional manner. In this paper
we present an extension of refinement calculus for reactive systems.
Refinement calculus is based on monotonic predicate transformers,
which transform sets of post-states into sets of pre-states. To model
reactive systems, we introduce monotonic property transformers, which
transform sets of output traces into sets of input traces. We show
how to model in this semantics refinement, sequential composition,
demonic choice, and other semantic operations on reactive systems.
We use primarily higher order logic to express our results, but we
also show how property transformers can be defined using other formalisms
more amenable to automation, such as linear temporal logic (suitable
for specifications) and symbolic transition systems (suitable for
implementations). Finally, we show how this framework generalizes
previous work on relational interfaces so as to be able to express
systems with infinite behaviors and liveness properties.
\end{abstract}

\section{Introduction}

Refinement calculus \cite{back-1978,back-wright-98} is a powerful
and expressive tool for reasoning about sequential programs. Refinement
calculus is based on a \emph{monotonic predicate transformer} semantics
which allows to model total correctness (functional correctness and
termination), unbounded nondeterminism, demonic and angelic nondeterminism,
among other program features. The framework also allows to express
compatibility during program composition (e.g., whether the postcondition
of a statement is strong enough to guarantee the precondition of another)
and also to reason about program evolution and substitution via refinement. 

As an illustrative example, consider a simple assignment statement
performing division: $z:=x/y$. Semantically, this statement is modeled
as a predicate transformer, denoted $\div$. $\div$ is a function
which takes as input a predicate $q$ characterizing a set of program
states and returns a new predicate $p$ such that if the program is
started in any state in $p$ it is guaranteed to terminate and reach
a state in $q$ (that is, $p$ is the \emph{weakest precondition}
of $q$). For our division example, we would also like to express
the fact that division by zero is not allowed. To achieve this, we
can define the predicate transformer as follows: $\div(q)=\{(x,\; y,\; z)\;|\; y\not=0\land(x,\; y,\; x/y)\in q\}$.

Having defined the semantics of the division statement, we can now
compose it with another statement, say, a statement that reads the
values of $x$ and $y$ from the console: $(x,y):=read()$. Making
no assumptions on what \emph{read} does, we model it as the so-called
$\havoc$ statement, which assigns arbitrary values to program variables.
Formally, \emph{read} is modeled as the predicate transformer: $\havoc(q)=\begin{cases}
\top & \mbox{if}\; q=\top\\
\bot & \mbox{otherwise}
\end{cases}$ where $\top$ and $\bot$ denote the universal and empty sets, respectively.
Now, what happens if we compose the two statements in sequence? That
is, $(x,y):=read();z:=x/y$. Refinement calculus tells us that sequential
composition of statements corresponds to function composition of their
predicate transformers, so the semantics of the composition is $\havoc\circ\div$,
which can be shown to be equivalent to the predicate transformer $\fail$,
defined as $\fail(q)=\bot$ for any $q$. This indicates incompatibility,
i.e., the fact that the composition of the two statements is invalid.
Indeed, without any assumptions on \emph{read}, we cannot guarantee
absence of division by zero.

We can go one step further and reason about program substitution via
refinement. Assume we have another division statement, but this time
it calculates only some approximation of the result: $z:=z'$ such
that $abs(x/y-z')\le\epsilon$. We model this new division statement
as a new predicate transformer $\div'$ defined as follows: $\div'(q)=\{(x,\; y,\; z)\;|\; y\not=0\land(\forall z':abs(x/y-z')\le\epsilon\Rightarrow(x,\; y,\; z)\in q)\}$.
Refinement calculus allows us to state and prove that $\div$ refines
$\div'$, and conclude that the $\div$ statement can substitute the
$\div'$ statement without affecting the properties of the overall
program.

Refinement calculus has been developed so far primarily for sequential
programs. In this paper we present an extension of refinement calculus
for \emph{reactive systems} \cite{Harel:1989:DRS:101969.101990}.
Denotationally, a reactive system can be seen as a system which accepts
as input infinite sequences of values, and produces as output infinite
sequences of values. Operationally, a reactive system can be seen
as a machine with input, output, and state variables, which operates
in steps, each step consisting of reading the inputs, writing the
outputs, and updating the state. Our framework allows us to specify
a very large class of reactive systems, including nondeterministic
and non-receptive systems, with both safety and liveness properties,
both denotationally and operationally. It also allows to define system
composition and to talk about incompatibility, refinement, and so
on. To illustrate these features, we provide an example analogous
to the division example above.
\begin{example}
\label{example_intro}Consider the two components shown in Figure
\ref{fig:incomp-1} and specified using the syntax of linear temporal
logic \cite{pnueli:1977}. Component $A=\always(x\ge0)$ specifies
that its output $x$ is never less than zero, while component $B=\always\event(x=1)$
requires that its input is infinitely often equal to one (the fact
that the output of $B$ has the same label $x$ as the input means
that $B$ sets its output to be equal to the input -- provided the
input requirement holds). The output of $A$ is connected to the input
of $B$. Using our framework, we can show that this composition is
invalid, that is, that $A$ and $B$ are incompatible, because the
output guarantee of $A$ is not strong enough to satisfy the input
requirement of $B$.

\vspace{-1ex}

\begin{figure}[H]
\begin{centering}
\includegraphics{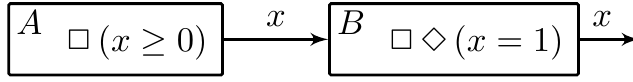}
\par\end{centering}

\vspace{-2ex}

\centering{}\caption{\label{fig:incomp-1}Two incompatible systems}
\end{figure}

\vspace{-4ex}

\begin{figure}[H]
\begin{centering}
\includegraphics{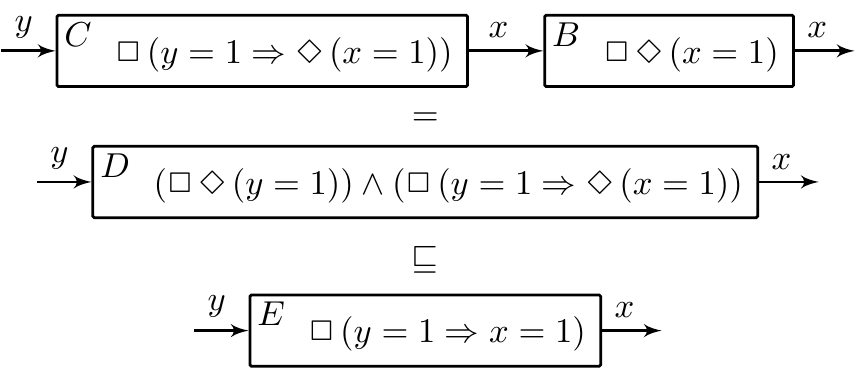}
\par\end{centering}

\centering{}\vspace{-2ex}
\caption{\label{fig:comp-1}Two compatible systems (top), their composition
(middle), and a refinement (bottom)}
\end{figure}

\vspace{-1ex}
The above is akin to behavioral type checking. We can also use our
framework to perform behavioral type inference. We can deduce, for
instance, that component $C=\always(y=1\Rightarrow\event(x=1))$ from
Figure \ref{fig:comp-1}, which models a request-response property
(always $y=1$ implies eventually $x=1$) is compatible with component
$B$ above, and infer automatically a new input requirement $\always\event(y=1)$
for the composite system $D$.

Finally, we can reason about refinement, akin to behavioral subtyping.
In the example of Figure \ref{fig:comp-1}, we can show that the executable
component $E$ which sets output $x=1$ whenever input $y=1$ refines
the component $D$, and therefore conclude that $E$ can substitute
$D$ in any context.
\end{example}
The key technical contribution of our paper, which allows us to develop
a refinement calculus of reactive systems, is the notion of \emph{monotonic
property transformers}. A property transformer is a function which
takes as input an \emph{output property} $q$ and returns an \emph{input
property} $p$. Properties are sets of traces, so that $q$ is a set
of output traces and $p$ is a set of input traces. In other words,
similarly to predicate transformers, which transform postconditions
to preconditions, property transformers transform \emph{out-conditions}
to \emph{in-conditions}.

Monotonic property transformers (MPTs) provide the semantical foundation
for system specification and implementation in our framework. We generally
use higher order logic to specify MPTs, but we also show how to MPTs
can be defined using formalisms more amenable to automation, such
as linear temporal logic and \emph{symbolic transition systems} (similar
to the formalism used by the popular model-checker NuSMV). We also
provide the basic operations on MPTs: composition, compatibility,
refinement, variable hiding, and so on. We study subclasses of MPTs
specified by input-output relations, and derive a number of interesting
closure and other properties on them. Finally, as an application of
our framework, we show how it can be used to extend the relational
interfaces framework of \cite{tripakis:2011} from only safety (finite,
prefix-closed) properties, to also infinite properties and liveness.

In the sequel we use higher order logic as implemented in Isabelle/HOL
\cite{nipkow-paulson-wenzel-02} to express our concepts. All results
presented in this paper were formalized in Isabelle, and our presentation
translates directly into Isabelle's formal language. The Isabelle
formalization is available from the Archive of Formal Proofs \textcolor{blue}{\cite{preoteasa:2014afp}.}

\subsection{Related work}

A number of compositional frameworks for the specification and verification
of input-output reactive systems have been proposed in the literature.
In the Focus framework \cite{BroyStolen01} specifications are relations
on input-output streams. Focus is able to express infinite streams
and liveness properties, however, it focuses on \emph{input-receptive}
systems, that is, systems where all input values are always legal.
Other compositional frameworks that also assume input-receptiveness
are Dill's \emph{trace theory} \cite{dill:1989}, \emph{IO automata}
\cite{LynchTuttle89}, and \emph{reactive modules} \cite{AlurHenzingerFMSD99}.
Our framework allows to specify non-input-receptive systems, where
some inputs are sometimes illegal. For example, in the case of the
division statement $[z:=x/y]$, we can write $y\neq0\land\cdots$
instead of $y\neq0\Rightarrow\cdots$. The conjunction specifies a
non-input-receptive system where $y=0$ is an illegal input, whereas
the implication specifies an input-receptive system. As argued in
\cite{tripakis:2011}, the ability to specify illegal inputs is essential
in order to obtain a lightweight verification framework, akin to type-checking.
In particular, it allows to define a behavioral notion of component
\emph{compatibility}, which goes beyond syntactic compatibility (correct
port matching) as illustrated by the examples given above.

There are also compositional frameworks which allow to specify non-input-receptive
systems. Among such frameworks, our work is inspired from refinement
calculus, on one hand, and \emph{interface theories} on the other,
such as \emph{interface automata} \cite{AlfaroHenzingerFSE01} and
\emph{relational interfaces} \cite{tripakis:2011}. These interface
theories, however, cannot express liveness properties. The same is
true with existing extensions of refinement calculus to infinite behaviors
such as \emph{action systems} \cite{back:1990,back:wright:1994},
which do not have acceptance conditions (say, of type B�chi) and therefore
cannot express general liveness properties. \emph{Fair} action systems
\cite{back:xu:1998}, augment action systems with fairness assumptions
on the actions, but it is unclear whether they can handle general
liveness properties, e.g., full LTL. Our approach is based on a natural
generalization from predicate to property transformers, and as such
can handle liveness (and more) as part of system specification.

\section{Preliminaries}

We use capital letters $X$, $Y$, $\Sigma$, $\ldots$ to denote
types, and small letters to denote elements of these types $x\in X,$....
We denote by $\bool$ the type of the Boolean values $\tru$ and $\fals$,
and by $\nat$ the type of natural numbers. We use in general the
sans-serif font to denote constants (types and elements). We use $\land$,
$\lor$, $\Rightarrow$, $\neg$ for the Boolean operations.

If $X$ and $Y$ are types, then $X\to Y$ denotes the type of functions
from $X$ to $Y$. We use a \emph{dot notation} for function application,
so we write $f.x$ instead of $f(x)$ from now on. If $f:X\to Y\to Z$
is a function which takes the first argument from $X$ and the second
argument from $Y$ and the result is from $Z$, and if $x\in X$ and
$y\in Y$ then $f.x.y$ denotes the function $f$ applied to $x$
and the result applied to $y$. According to this interpretation function
application associates to the left ($f.x.y=(f.x).y$) and correspondingly
the function type constructor ($\to$) associates to the right ($X\to Y\to Z=X\to(Y\to Z)$).
We use also lambda notation for constructing functions. For example
if $x+y+2\in\nat$ is a natural expression then $(\lambda x,\; y:x+y+2):\nat\to\nat$
is the function which maps $x$ and $y$ to $x+y+2$. We use the notation
$X\times Y$ for the Cartesian product of $X$ and $Y$, and if $x\in X$
and $y\in Y$, then $(x,\, y)$ is a pair from $X\times Y$. 

Predicates are functions returning Boolean values (e.g., $p:X\to Y\to\bool$),
and relations are predicates with at least two arguments. For a relation
$r:X\to Y\to\bool$ we denote by $\inp.r:X\to\bool$ the predicate
given by
\[
\inp.r=(\exists y:r.x.y)
\]
If $r$ is a relation with more than two arguments then we define
$\inp.r$ similarly by quantifying over the last argument of $r$:
\[
\inp.r.x.y.z=(\exists u:r.x.y.z.u)
\]
We extend point-wise the operations on $\bool$ to operations on predicates.
For example, if $p$ is a predicate only on $x$, i.e., $p:X\to\bool$
and $q$ is a predicate on $x$ and $y$, i.e., $q:X\to Y\to\bool$,
then:
\[
(p\land q).x.y=p.x\land q.x.y
\]
and we also have in this case: 
\[
p\land(\inp.q)=\inp.(p\land q)
\]

We use $\bot$ and $\top$ as the smallest and greatest predicates
\[
\bot.x=\fals\mbox{ and }\top.x=\tru
\]

The composition of relations $r,\; r'$ is denoted $r\circ r'$ and
it is a relation given by: 
\[
(r\circ r').x.z=(\exists y:r.x.y\land r'.y.z)
\]
We treat subsets of a type, and predicates with one argument as being
the same and we use both notations $x\in p$ and $p.x$ to express
the fact that $p$ is true in $x$. For constructing predicates we
use lambda abstraction (e.g., $\lambda x,\; y:x\le10\Rightarrow y>10$),
and for predicates with single arguments we use also set comprehension
$\{x\;|\; x>10\}$.

We assume that $\Sigma$ is a type of program states. For example
for imperative programs over some variables $x$, $y$, $z$, $\ldots$,
a state $s\in\Sigma$ gives values to the program variables $x,\; y,\; z,\;\ldots$.
In general, the systems that we consider may have different input
and output variables, and we can also have different state sets. For
a system with a variable $x$, $\Sigma_{x}$ denotes the type of states
which gives values to $x$. For a state $s\in\Sigma$, $x.s$ is the
value of $x$ in $s$ and $s[x:=a]$ is new state obtained from $s$
by changing the value of $x$ to $a$.

For reactive systems we model states as \emph{infinite sequences}
or \emph{traces} from $\Sigma$. Formally such an infinite sequence
is an element $\sigma\in\Sigma^{\omega}$ where $\Sigma^{\omega}=(\nat\to\Sigma)$.
For $\sigma\in\Sigma^{\omega}$, $\sigma_{i}=\sigma.i$, and $\sigma^{i}\in\Sigma^{\omega}$
is given by $\sigma_{j}^{i}=\sigma^{i}.j=\sigma_{i+j}$. We consider
a pair of traces $(\sigma,\sigma')$ as being the same as a trace
of pairs $(\lambda i:(\sigma_{i},\sigma'_{i}))$. 

In the next subsection we introduce \emph{linear temporal logic} (LTL)
which is the main logic that we use to specify reactive systems.

\subsection{Linear temporal logic}

Linear temporal logic (LTL) \cite{pnueli:1977} is a logic used for
specifying properties of reactive systems. In addition to the connectives
of classical logic it contains modal operators referring to time.
LTL formulas can express temporal properties like something is always
true, or something is eventually true, and their truth values are
given for infinite sequences of states. For example the formula $\always x=1$
(always $x$ is equal to 1) is true for the infinite sequence $\sigma$
if for all $i\in\nat$ $x.\sigma_{i}=1$.

The semantics of an LTL formula is the set of all sequences for which
the formula is true. In this paper we use a semantic (algebraic) version
of LTL. For us an LTL formula is a predicate on traces and the temporal
operators are functions mapping predicates to predicates. We call
predicates over traces (i.e., sets of traces) \emph{properties}. 

If $p,\; q\in\Sigma^{\omega}\to\bool$ are properties, then \emph{always}
$p$, \emph{eventually} $p$, \emph{next} $p$, and $p$ \emph{until}
$q$ are also properties and they are denoted by $\always p$, $\event p$,
$\nex p$, and $p\until q$ respectively. The property $\always p$
is true in $\sigma$ if $p$ is true at all time points in $\sigma$,
$\event p$ is true in $\sigma$ if $p$ is true at some time point
in $\sigma$, $\nex p$ is true in $\sigma$ if $p$ is true at the
next time point in $\sigma$, and $p\until q$ is true in $\sigma$
if there is some time in $\sigma$ when $q$ is true, and until then
$p$ is true. Formally we have:
\[
\begin{array}{lll}
(\always p).\sigma & = & (\forall n:p.\sigma^{n})\\
(\event p).\sigma & = & (\exists n:p.\sigma^{n})\\
(\nex p).\sigma & = & p.\nxt{\sigma}\\
(p\until q).\sigma & = & (\exists n:(\forall i<n:p.\sigma^{i})\land q.\sigma^{n})
\end{array}
\]

Quantification for properties is defined in the following way
\[
(\forall x:p).\sigma=(\forall a:p.(\sigma[x:=a]))
\]
where $a$ ranges over infinite traces of $x$ values, and $\sigma[x:=a].i=\sigma_{i}[x:=a_{i}]$.
When $p$ is a predicate on traces $x$ and $y$, then quantification
is defined as normally in predicate calculus, as in $\forall a:p.a.b$. 

We lift normal arithmetic and logical operations to traces ($x$ and
$y$) in the following way
\[
\begin{array}{lll}
x+y & = & x_{0}+y_{0}\\
x\land y & = & x_{0}\land y_{0}
\end{array}
\]

\begin{lem}
If $p$ and $q$ are properties, then we have: $(\exists x:\always p)=\always(\exists x:p)$
and $\always(\inp.p)=\inp.(\always p)$.\end{lem}
\begin{defn}
We define the operator $p\leads q=\neg(p\until\neg q)$. Intuitively,
$p\leads q$ holds if, whenever $p$ holds continuously up to step
$n-1$, then $q$ must hold at step $n$.\end{defn}
\begin{lem}
If $p$ and $q$ are properties, then we have\end{lem}
\begin{enumerate}
\item $(p\leads q).\sigma=(\forall n:(\forall i<n:p.\sigma^{i})\Rightarrow q.\sigma^{n})$
\item $p\leads p=\always p$ and $\tru\leads p=\always p$
\end{enumerate}
Using LTL properties we can express \emph{safety} properties, expressing
that \emph{something bad never happens} (e.g., $\always t\le10{}^{\circ}$
-- the temperature stays always below $10^{\circ}$), as well as \emph{liveness}
properties, expressing that \emph{something good eventually happens
}(e.g., $\always\event x=0$ -- infinitely often $x$ becomes $0$).

\section{Monotonic property transformers}

\emph{Monotonic predicate transformers} are a powerful formalism for
modeling programs. A program $S$ from state space $\Sigma_{1}$ to
state space $\Sigma_{2}$ is formally modeled as a monotonic predicate
transformer, that is, a monotonic function from $(\Sigma_{2}\to\bool)\to(\Sigma_{1}\to\bool)$,
with a weakest precondition interpretation. If $S$ is a program and
$q\in\Sigma_{2}\to\bool$ is a predicate on $\Sigma_{2}$ (set of
final states), then $S.q$ is the set of all initial states from which
the execution of $S$ always terminates and it terminates in a state
from $q$. \emph{Monotonic Boolean transformers} (MBTs) \cite{preoteasa:2013}
is a generalization of monotonic predicate transformers, where instead
of predicates $(\Sigma_{i}\to\bool)$ arbitrary complete Boolean algebras
are used. MBTs are monotonic functions from a complete Boolean algebra
$B_{2}$ to a complete Boolean algebra $B_{1}$. 

In this section we introduce \emph{monotonic property transformers}
(MPTs), and we use them to model input-output reactive systems. MPTs
are MBTs from the complete Boolean algebra of $\Sigma_{y}$ properties
($\Sigma_{y}^{\omega}\to\bool$) to the complete Boolean algebra of
$\Sigma_{x}$ properties ($\Sigma_{x}^{\omega}\to\bool$), where $x$
and $y$ are the input and output variables, respectively. If $S$
is a reactive system with input variable $x$ and output variable
$y$, then a \emph{legal execution} of $S$ takes as input a sequence
of values for $x$, $\sigma=x_{0},\; x_{1},\;\ldots$, and produces
a sequence of values for $y$, $\sigma'=y_{0},\; y_{1},\;\ldots$.
This execution may be nondeterministic, that is, for the same input
sequence $\sigma$ we can obtain different output sequences $\sigma'$.
The execution of $S$ from $\sigma$ may also \emph{fail} if $\sigma$
does not satisfy some requirements on the input variables. As a property
transformer, the system $S$ is applied to a property $q\in\Sigma_{y}^{\omega}\to\bool$,
i.e., to a set of sequences over the output variable $y$. Then, $S$
returns the set of all sequences over the input variable $x$ from
which all executions of $S$ do not fail and produce sequences in
$q$. 

$S$ must be monotonic in the following sense: interpreting properties
as sets, $S$ is \emph{monotonic} if for any two properties $q,\ q'$,
if $q\subseteq q'$ then $S.q\subseteq S.q'$.

\subsection{Property transformers based on LTL}

Monotonic property transformers are appropriate primarily as semantic
descriptions of systems. In practice, we also need some \emph{syntax}
for describing systems in general, and property transformers in particular.
In this paper, we use two types of syntax: LTL, and \emph{symbolic
transition systems}. Property transformers based on symbolic transition
systems will be discussed in detail in Section \ref{sec_STS}. Property
transformers based on LTL are a special case of \emph{relational property
transformers}, discussed in detail in Section \ref{sec_Relational-property-transformers}.
Here we provide an illustrative example. 
\begin{example}
\label{ex_first_property_transf}Consider again component $B$ from
Example \ref{example_intro}, Figure \ref{fig:incomp-1}. Suppose
variable $x$ is a Boolean, taking values in the set $\{0,1\}$, i.e.,
$\Sigma_{x}=\{0,1\}$. Then, $B$ can be modeled as a property transformer
which from the set of properties $(\{0,1\}^{\omega}\to\bool)$ to
the same set (because $B$ copies its input to its output, provided
the requirements on the input are satisfied). Let $q\subseteq\{0,1\}^{\omega}$.
Then $B.q$ must contain exactly those infinite input sequences $\sigma\in\{0,1\}^{\omega}$
such that: (1) $\sigma$ satisfies the input requirement expressed
by the LTL property $\always\event(x=1)$, i.e., $\sigma$ must contain
infinitely many 1's; and (2) $\sigma$ is in $q$, since $B$ copies
its input to its output. Written formally, $B.q=\{\sigma\in q\mid(\always\event x=1).\sigma\}$.
Clearly, $B$ is a monotonic property transformer, as the larger the
set $q$ is, the larger $B.q$ is.
\end{example}

\subsection{Using property transformers as implicit system specifications}
\begin{example}
\label{ex_property_transformers} Example \ref{ex_first_property_transf}
provided the explicit definition of the property transformer for a
certain system, thereby also essentially completely defining that
system. Using property transformers, we can also specify systems \emph{implicitly},
by imposing constraints that the property transformers of these systems
must satisfy. In this way, we can specify the fact that a certain
system must exhibit various properties that we are interested in.
For example, the specification of a system $S$ that guaranties the
\emph{liveness} property that the output Boolean variable $y$ is
true infinitely often regardless of the input, is given by
\[
S.\{y\,|\,\always\event y\}=\top
\]

Note that the above equation does not define $S$ completely, it only
specifies a constraint that $S$ (interpreted as a property transformer)
must satisfy. Below, in Section \ref{sub:Basic-operations-on} we
give a complete definition of a MPT which satisfies the requirement
above.

Similarly, the specification of a system $S'$ that guaranties the
liveness property that the output Boolean variable $y$ is true infinitely
often when the integer input variable $x$ is equal to one infinitely
often, is given by
\[
\{x\,|\,\always\event x=1\}\subseteq S'.\{y\,|\,\always\event y\}
\]

\end{example}

\subsection{\label{sub:Basic-operations-on}Basic operations on monotonic property
transformers}

The point-wise extension of the Boolean operations to properties,
and then to monotonic property transformers gives us a \emph{complete
lattice} with $\sqsubseteq$ as the lattice \emph{order}, $\sqcap$
as the \emph{greatest lower bound}, or \emph{meet}, $\sqcup$ as the
\emph{least upper bound}, or \emph{join}, $\fail$ as the \emph{bottom}
element, and $\magic$ as the \emph{top} element. If $S$ and $T$
are monotonic property transformers, and $q$ is a property, then
these elements are formally defined by

\[
\begin{array}{lll}
(S\sqsubseteq T) & = & (\forall q:S.q\subseteq T.q)\\
(S\sqcap T).q & = & S.q\cap T.q\\
(S\sqcup T).q & = & S.q\cup T.q\\
\fail.q & = & \bot\\
\magic.q & = & \top
\end{array}
\]
Note that $\sqcap$ and $\sqcup$ preserve monotonicity. Also note
that, for any $S$, $\fail\sqsubseteq S$ and $S\sqsubseteq\magic$,
so indeed $\fail$ and $\magic$ are the bottom and top elements,
respectively. The transformer $\fail$ does not guarantee any property.
For any property $q$, we have $\fail.q=\bot$, i.e., there is no
input sequence for which $\fail$ will produce an output sequence
from $q$. On the other hand $\magic$ can establish any property
$q$ (for any $q,$ $\magic.q=\top$). The problem with $\magic$
is that it cannot be implemented.

All these lattice operations are also meaningful as operations on
reactive systems. The order of this lattice ($S\sqsubseteq T$) gives
the \emph{refinement relation} of reactive systems. If $S\sqsubseteq T$,
then we say that $T$ \emph{refines} $S$, or $S$ is \emph{refined}
\emph{by} $T$. If $T$ refines $S$ then we can replace $S$ with
$T$ in any context. Note that in some works (e.g., \cite{AlfaroHenzingerFSE01,tripakis:2011})
the notation is inverted, with $S\sqsubseteq T$ denoting $S$ refines
$T$, instead of $T$ refines $S$ as we employ here. In this paper
we follow the same convention as in \cite{back-wright-98}, which
is also consistent with the definition of refinement for property
transformers: $S\sqsubseteq T$ iff $S.q$ is a subset of $T.q$.

The interpretation of the lattice order as refinement follows from
the modeling of reactive systems as monotonic property transformers.
For example if we assume that $S$ and $S'$ introduced in Example
\ref{ex_property_transformers} are completely defined by
\[
S.q=\begin{cases}
\top & \mbox{ if }\{y\,|\,\always\event y\}\subseteq q\\
\bot & \mbox{ otherwise}
\end{cases}
\]
 and
\[
S'.q=\begin{cases}
\{x\,|\,\always\event x=1\} & \mbox{ if }\{y\,|\,\always\event y\}\subseteq q\\
\bot & \mbox{ otherwise}
\end{cases}
\]
then $S$ and $S'$ are monotonic and $S$ refines $S'$ ($S'\sqsubseteq S$).
In this example we see that if $S'$ is used within some context where
for certain inputs it guaranties outputs where $y$ is true infinitely
often, then $S$ can replace $S'$ because $S$ guaranties the same
property of the output regardless of its input.

The operations $\sqcap$ and $\sqcup$ model (\emph{unbounded}) \emph{demonic}
and \emph{angelic} \emph{nondeterminism} or \emph{choice}. The interpretation
of the demonic choice is that the system $S\sqcap T$ is correct (i.e.,
satisfies its specification) if both $S$ and $T$ are correct. In
this choice someone else (the demon) can choose to execute $S$ or
$T$, so they must both be correct. On the other hand the angelic
choice $S\sqcup T$ is correct if one of the systems $S$ and $T$
are correct. In this choice we have the control over the choice, and
we assume that we always choose the correct alternative. Unbounded
nondeterminism means that we could have unbounded choices as for example
in $\bigsqcap_{i\in I}S_{i}$ where $I$ is infinite. For example,
assume that we have two systems $S$ and $S'$ which compute the factorial
of $n$, but $S$ computes the correct result only for $n\le20$ and
$S'$ computes the correct result only for $10\le n$. Formally we
have
\[
\begin{array}{lll}
S.\{x\,|\, x=n!\} & = & \{n\,|\, n\le20\}\\
S'.\{x\,|\, x=n!\} & = & \{n\,|\,10\le n\}
\end{array}
\]
The demonic choice of $S$ and $S'$ is capable of computing the factorial
only for numbers between $10$ and $20$, while the angelic choice
will compute the factorial for all natural numbers $n$. 

\emph{Sequential composition} of two systems $S$ and $T$ is simply
the functional composition of $S$ and $T$ viewed as property transformers
($S\circ T$). We denote this type of composition by $S\comp T$ ($(S\comp T).q=S.(T.q)$).
To be able to compose $S$ and $T$, the type of the output of $S$
must be the same as the type of the input of $T$. 

The system $\skip$ defined by $(\forall q:\skip.q=q)$ is the neutral
element for sequential composition:
\[
\skip\comp S=S\comp\skip=S,\mbox{ for any }S.
\]
It is easy to see that sequential composition preserves monotonicity,
that is, if $S$ and $T$ are both monotonic property transformers,
then so is $S\comp T$.
\begin{defn}
Two systems $S$ and $T$ are \emph{incompatible} (w.r.t. the sequential
composition $S\comp T$) if 
\[
S\comp T=\fail.
\]

\end{defn}
Intuitively, $S$ and $T$ are compatible if the outputs of $S$ can
be controlled so that they are legal inputs for $T$. Controlling
the outputs of $S$ might mean restricting its own legal inputs. 
\begin{example}
\label{ex_compatibility}If for example we have $S$ and $T$ given
by
\[
S.q=\begin{cases}
\top & \mbox{ if }\{x\,|\, x>5\}\subseteq q\\
\bot & \mbox{ otherwise}
\end{cases}\mbox{ and }T.q=\{x\,|\, x<10\}
\]
then $(S\comp T).q=S.(T.q)=S.\{x\,|\, x<10\}=\bot$, for any $q$.
Therefore, $S$ and $T$ are in this case incompatible. This is because
$T$ requires its input to be smaller than 10, but $S$ can only guarantee
that its output will be greater than 5, and there is no way to restrict
the input of $S$ to make this guarantee stronger.

On the other hand, assuming that the input and output of $S$ and
$T$ have the same type, $T$ and $S$ are compatible w.r.t. the reverse
composition, i.e., $T\comp S$ is not $\fail$. Indeed, we have $T.(S.q)=\{x\,|\, x<10\}$,
for any $q$. \end{example}
\begin{defn}
For a property transformer $S$, the \emph{fail }of $S$, denoted
$\pfail.S$, is the set of \emph{illegal} input sequences, i.e., the
set of input sequences for which the system produces no output, or
\emph{fails} to establish any output property. Formally:
\[
\pfail.S=\neg S.\top.
\]

\end{defn}
For example, the fail of $\magic$ is $\bot$ and the fail of $\fail$
is $\top$.
\begin{defn}
For a property transformer $S$, the \emph{guard} of $S$, denoted
$\grd.S$, is the set of input sequences for which the system does
not behave \emph{miraculously}. Formally:
\[
\grd.S=\neg S.\bot.
\]

\end{defn}
For example, the guard of $\magic$ is $\bot$ and the guard of $\fail$
is $\top$. To see the intuition behind the definition of guard, observe
that $S.\bot$ is the set of input sequences for which $S$ is guaranteed
to establish $\bot$, that is, the empty property, and therefore by
monotonicity of $S$ also any other output property. In other words,
$S.\bot$ is the set of inputs for which $S$ behaves miraculously,
since the empty property $\bot$ cannot be established. 

For instance, taking $S$ and $T$ to be as defined in Example \ref{ex_compatibility},
we have: $\pfail.S=\bot$, $\grd.S=\top$, and $\pfail.T=\grd.T=\{x\mid x\ge10\}.$

\subsection{Assert and demonic update transformers}

We now define two special types of property transformers which will
be used to form more general property transformers by composition.
For $p,\; q\in\Sigma^{\omega}\to\bool$ and $r\in\Sigma_{1}^{\omega}\to\Sigma{}_{2}^{\omega}\to\bool$
we define the \emph{assert property transformer} $\assert p:(\Sigma^{\omega}\to\bool)\to(\Sigma^{\omega}\to\bool)$,
and the \emph{demonic update property transformer} $\assume r:(\Sigma_{2}^{\omega}\to\bool)\to(\Sigma_{1}^{\omega}\to\bool)$
as follows:
\[
\begin{array}{ll}
\assert p.q & =p\cap q\\
\assume r.q.\sigma & =(\forall\sigma':r.\sigma.\sigma'\Rightarrow q.\sigma')
\end{array}
\]
The assert transformer $\assert p$ models a system which, given input
sequence $\sigma$, produces $\sigma$ as output when $p.\sigma$
is true, and it fails otherwise. In other words, only inputs satisfying
$p$ are legal for the assert system. The demonic update transformer
$\assume r$ models a system which establishes a post condition $q$
when given as input a sequence $\sigma$ if all sequences $\sigma'$
with $r.\sigma.\sigma'$ are in $q$. Note that the assert and demonic
update property transformers are monotonic, for any $p$ and $r$.
\begin{example}
The property transformer for component $B$ of Figure \ref{fig:incomp-1},
discussed already in Example \ref{ex_first_property_transf}, is an
example of an assert property transformer $\{p\}$, where $p.\sigma$
holds iff $\sigma$ satisfies the LTL formula $\always\event(x=1)$.
Component $E$ of Figure \ref{fig:comp-1} is an example of a demonic
update property transformer $[r]$, where $r$ is the input-output
trace relation corresponding to the LTL formula $\always(y=1\Rightarrow x=1)$. 
\end{example}

\subsection{Notation for assert and demonic update}

Let us now introduce some preliminary syntactic notations to describe
the two kinds of property transformers introduced above. Let $R$
be an expression in $x$ and $y$, for example, the LTL formula $\always(x=1\Rightarrow y=1)$.
Recall that $\lambda x,y:R$ denotes the function $r:\Sigma_{x}^{\omega}\to\Sigma{}_{y}^{\omega}\to\bool$
that takes two sequences $x$ and $y$ and returns true iff these
two sequences satisfy the LTL formula. Since $r$ is also an input-output
relation on sequences, it defines the demonic property transformer
$[r].$ However, a notation such as $[\lambda x,y:\always(x=1\Rightarrow y=1)]$
may be heavier than necessary. Therefore, we introduce a lighter notation,
namely, $\demonic xy{\always(x=1\Rightarrow y=1)}$. In general, for
any expression $R$ in $x$ and $y$, we use notation $\demonic xyR$
as equivalent to $\assume{\lambda x,y:R}$. This notation also extends
to systems with more than one inputs or outputs. For example, if $R$
is $z=x+y$, and $x,y$ are the inputs while $z$ is the output, then
$\demonic{x,y}z{z=x+y}=\assume{\lambda(x,y),z:z=x+y}$. 

For assert property transformers we introduce similar lighter notation.
If $P$ is an expression in $x$ then $\assert{x\;|\; P}=\assert{\lambda x:P}$.
For example, if $P$ is $x\le y$, then $\assert{x,y\;|\; x\le y}=\assert{\lambda(x,y):x\le y}$.
Note that a notation such as $\{x\mid x<1\}$ is ambiguous, as it
could mean the set of all, say, real numbers smaller than 1, or the
assert property transformer $\{\lambda x:x<1\}$. We will still use
such notation, however, and such ambiguity will be resolved from the
context. 

Note also that in notations such as $\assert{x\;|\; P}$ and $\demonic xyR$
, the variables $x$ and $y$ are bound. However, when we compose
some of these property transformers we will try whenever possible
to use the same name for the output variables of a transformer which
are input to another transformer. For example, we will use the notation:
\[
\assert{x,y\;|\; x\le y}\comp\demonic{x,y}z{z=x+y}\comp\demonic zu{u=z^{2}}
\]
instead of the equivalent one:
\[
\assert{x,y\;|\; x\le y}\comp\demonic{u,v}x{x=u+v}\comp\demonic ux{x=u^{2}}.
\]
Sometimes we also need demonic transformers that copy some of the
input variables into some of the output variables, as in, for example
\[
S=\demonic{u,x}{y,v}{(x=y)\land r.u.x.y.v}.
\]
In this case, we drop the condition $x=y$ from the relation of $S$
and we simply use the same name for $x$ and $y$:
\[
S=\demonic{u,x}{x,v}{r.u.x.x.v}.
\]
If we want to rearrange the input variables into the output variables
and if we want to drop some input variables and introduce some new
arbitrary variables, then we use syntax like the following: 
\[
S=\assume{x,y,u,z,x\leadsto z,y,x,y,v}
\]
This notation stands for 
\[
S=[x,y,u,z,x'\leadsto z',y',x'',y'',v\mid x=x'=x''\land y=y'=y''\land z=z']
\]
which is equivalent to
\[
S=[\lambda(x,y,u,z,x'),\;(z',y',x'',y'',v):x=x'=x''\land y=y'=y''\land z=z']
\]
If $S$ starts on a tuple where the first component is the same as
the last component ($x=x'$), then $S$ returns $z',\, y',\, x'',\, y'',\, v$
such that $x=x'=x''\land y=y'=y''\land z=z'$. On the other hand if
$S$ starts on a tuple where the first component is different from
the last component, then $S$ behaves miraculously.

\subsection{Properties of assert and demonic update}
\begin{thm}
If $p,q\in\Sigma^{\omega}\to\bool$, $r\in\Sigma_{1}^{\omega}\to\Sigma{}_{2}^{\omega}\to\bool$,
and $r'\in\Sigma_{2}^{\omega}\to\Sigma{}_{3}^{\omega}\to\bool$, then\end{thm}
\begin{enumerate}
\item \label{enu:assert-dem-a-1}$\skip=\assume{x\leadsto x}=\assert{x\mid\tru}$
($\skip$ is both a demonic update and an assert transformer)
\item $\magic=\assume{x\leadsto y\mid\fals}$, and $\fail=\assert{x\mid\fals}$
($\magic$ is a demonic update, and $\fail$ is an assert transformer)
\item \label{enu:assert-dem-b-1}$\assert p\comp\assert{p'}=\assert{p\cap p'}$
and $\{x\mid P\}\comp\{x\mid P'\}=\{x\mid P\land P'\}$ (Assert transformers
are closed under sequential composition)
\item $\assume r\comp\assume{r'}=\assume{r\circ r'}$ and $[x\leadsto y\mid R]\comp[y\leadsto z\mid R']=[x\leadsto z\mid\exists y:R\land R']$
(Demonic updates are closed under sequential composition)
\item \label{enu:assert-dem-f-1}$\grd.\assert p=\top$, and $\grd.\assume r=\inp.r$
(Calculating the gard of assert and domonic update transformers)
\item $\pfail.\{p\}=\neg p$ and $\pfail.[r]=\bot$ (Calculating the fail
of assert and domonic update transformers)
\end{enumerate}

\section{Relational property transformers\label{sec_Relational-property-transformers}}
\begin{defn}
A \emph{relational property transformer} (RPT) is a property transformer
of the form $\assert p\comp\assume r$. The assert transformer $\assert p$
imposes the restriction $p$ on the input sequences, and the demonic
update $\assume r$ nondeterministically chooses output sequences
according to the relation $r$. For a RPT $S=\assert p\comp\assume r$
we call $p$ the \emph{precondition} of $S$ and $r$ the\emph{ input-output
relation} of $S$. For a RPT $\assert p\comp\assume r$ we use the
notation $\system pr$.
\end{defn}
For RPTs we introduce also syntactic notation similar to the one introduced
for assert and demonic transformers:
\begin{eqnarray*}
\system{x\leadsto y\;|\; P}R & = & \assert{x\;|\; P}\comp\demonic xyR
\end{eqnarray*}

Note that every assert property transformer $\{p\}$ is a relational
property transformer, because $\{p\}=\{p\}\comp[x\leadsto y\mid x=y]$.
Also, every demonic update property transformer is a relational property
transformer, because $[r]=\{x\mid\tru\}\comp[r]$. Also note that
every relational property transformer is by definition monotonic.
This is because every assert transformer is monotonic, every demonic
update transformer is monotonic, and monotonicity is preserved by
sequential composition. Finally, note that, as a special case, property
$p$ and relation $r$ can be described by LTL formulas. This allows
us to describe RPTs syntactically, by means of LTL formulas. This
is illustrated in the example that follows.
\begin{example}
\label{ex_division_transformers}Consider again the division statement
$z:=x/y$ discussed in the introduction. Using LTL and the syntax
introduced above, we can define several variants of property transformers
which perform division on sequences of input pairs $x$ and $y$,
as follows:
\[
\begin{array}{lllc}
S_{1} & = & [x,y\leadsto z\mid\always(y\ne0\land z=x/y)]\\
S_{2} & = & \{x,y\leadsto z\mid\always y\ne0\mid\always(y\ne0\land z=x/y)] & =\{x,y\mid\always y\ne0\}\comp S_{1}
\end{array}
\]
$S_{1}$ and $S_{2}$ are different property transformers. Both are
relational, but they have different guards and fails. Specifically,
$\pfail.S_{1}=\bot$, whereas $\pfail.S_{2}=(\event y=0)$. This means
that any input trace is legal for $S_{1}$ whereas only traces where
$y$ is never zero are legal for $S_{2}$. On the other hand, $\grd.S_{1}=(\always y\ne0)$,
whereas $\grd.S_{2}=\top$. This means that $S_{2}$ never behaves
miraculously, whereas $S_{1}$ behaves miraculously when the input
assumption $\always y\ne0$ is violated. 
\end{example}
The next theorem\textcolor{red}{{} }states some important results, in
particular regarding the compositionality (i.e., closure w.r.t. composition
and other operations) of relational property transformers.
\begin{thm}
\label{lem:assert-demonic}Let $p,q$ be properties and $r,r'$ be
relations on sequences of appropriate types. Then:\end{thm}
\begin{enumerate}
\item $\system pr\comp\system{p'}{r'}=\system{x\leadsto z\;|\; p.x\land(\forall y:r.x.y\Rightarrow p'.y)}{(r\circ r').x.z}$
(relational property transformers are closed under sequential composition) 
\item \label{enu:assert-dem-d}$\system pr\sqcap\system{p'}{r'}=\system{p\land p'}{r\lor r'}$
(relational property transformers are closed under demonic choice)
\item \label{enu:assert-dem-c}$\system pr=\system p{p\land r}$ (precondition
can be used in the input-output relation, e.g., for simplification)
\item \label{enu:assert-dem-e}$\system pr\sqsubseteq\system{p'}{r'}\Leftrightarrow(\forall x:p.x\Rightarrow p'.x)\land(\forall x,y:(p.x\land r'.x.y)\Rightarrow r.x.y)$
(necessary and sufficient condition for refinement)
\item \label{enu:assert-dem-g}$\grd.(\system pr)=\neg p\lor\inp.r$ (symbolic
expression for the guard)
\item $\pfail.(\system pr)=\neg p$ (symbolic expression for the fail predicate)
\end{enumerate}
\textcolor{red}{}

\subsection{\label{sub:Guarded-systems}Guarded systems}

Relational property transformers are a strict subclass of monotonic
property transformers, but they still allow to describe systems that
may behave \emph{miraculously}. An example of a transformer that may
behave miraculously is transformer $S_{1}$ defined in Example \ref{ex_division_transformers}.
Often we are interested in systems that are guaranteed to never behave
miraculously, i.e., in systems defined by transformers $S$ such that
$\grd.S=\top$. In these cases we use relational property transformers
of the form $\system{\inp.r}r$. We call these RPTs \emph{guarded}:
\begin{defn}
The \emph{guarded system} of a relation $r$ is the relational property
transformer $\gsystem r=\system{\inp.r}r$.
\end{defn}
For guarded systems we also introduce a notation similar to the notation
introduced for relational property transformers:
\[
\gsystem{x\leadsto y\;|\; R}=\system{x\leadsto y\;|\;\inp.R}R.
\]

It is worth pointing out that the property transformer $\system{\inp.r}r$
is as general as $\system{p\land\inp.r}r$ because we have $\system{p\land\inp.r}r=\system{\inp.(p\land r)}{p\land r}$:
\begin{lyxlist}{00.00.0000}
\item [{~}] $\system{p\land\inp.r}r$
\item [{$=$}] \{Theorem \ref{lem:assert-demonic}\}
\item [{~}] $\system{p\land\inp.r}{p\land\inp.r\land r}$
\item [{$=$}] \{Theorem \ref{lem:assert-demonic}\}
\item [{~}] $\system{\inp.(p\land r)}{p\land r}$
\end{lyxlist}
The theorem that follows states several important closure properties
for guarded systems.
\begin{thm}
\label{thm:grd-sysnew}If $p$ is a property and $r,r'$ are relations
of appropriate types, then\end{thm}
\begin{enumerate}
\item $\grd.\gsystem r=\top$ (guarded systems never behave miraculously)
\item $\fail=\gsystem{\bot}$ and $\skip=\gsystem{x\leadsto x\;|\;\top}$
($\fail$ and $\skip$ are guarded)
\item $\assert p=\gsystem{x\leadsto x\;|\; p.x}$ and $\assert p\comp\gsystem r=\gsystem{p\land r}$
(assert transformers are guarded and assert can be moved inside a
guarded transformer)
\item $\gsystem r\comp\gsystem{r'}=\gsystem{x\leadsto z\;|\;\inp.r.x\land(\forall y:r.x.y\Rightarrow\inp.r'.y)\land(r\circ r').x.z}$
(guarded systems are closed under sequential composition)
\item $\gsystem r\sqcap\gsystem{r'}=\gsystem{\inp.r\land\inp.r'\land(r\lor r')}$
(guarded systems are closed under demonic choice)
\end{enumerate}
Note that part 3 of the above lemma implies that assert transformers
are special cases of guarded systems. However, a demonic update is
generally not a guarded system. For instance, we have $\grd.\assume{\bot}=\bot$.
A less pathological example is the demonic update transformer $S_{1}$
from Example \ref{ex_division_transformers}, which is also not a
guarded system, because it behaves miraculously when $y$ becomes
0. As the following lemma states, demonic updates are guarded systems
if and only if they impose no requirements on the inputs.
\begin{lem}
\label{lem:grd-sys-dem}The demonic update transformer $\assume r$
is a guarded system if and only if $\inp.r=\tru$ and in this case
we have $\assume r=\gsystem r$.\end{lem}
\begin{example}
\label{ex_guarded}Here are some examples of guarded systems:\end{example}
\begin{itemize}
\item $\havoc=\demonic xy{\tru}$: this demonic update transformer corresponds
to a system which accepts any input sequence, and may generate an
arbitrary output sequence. $\havoc$ is a guarded system because it
imposes no requirements on its input.
\item $\assertlive=\assert{x\;|\;\always(\event x)}$: this assert transformer
corresponds to a system which requires its Boolean input to be infinitely
often true.
\item $\livehavoc=\assertlive\comp\havoc$: this system corresponds to the
sequential composition of the previous two; it requires the input
to be infinitely often true, and it makes no guarantees on the output
(i.e., it can generate any output sequence).
\item $\reqresp=\demonic xy{\always(x\Rightarrow\event y)}$: this demonic
update transformer corresponds to a system which accepts any input
sequence, and may generate an arbitrary output sequence, provided
the request-response property \emph{for every input there is eventually
an output} is satisfied.
\end{itemize}
The fact that all these systems are guarded follows from Theorem \ref{thm:grd-sysnew}
and Lemma \ref{lem:grd-sys-dem}. Note that $\reqresp$ illustrates
the ability of our framework to express unbounded nondeterminism since,
for a given input sequence $x$, there is an infinite set of $y$
sequences that satisfy the request-response LTL formula. (We can also
express unbounded nondeterminism for systems with infinite data types.) 

For the above systems we have the following properties:
\begin{itemize}
\item $\havoc\comp\assertlive=\fail$: this means that $\havoc$ and $\assertlive$
are incompatible. Indeed, since $\havoc$ guarantees nothing about
its output, it cannot meet the input requirements of $\assertlive$.
\item $\havoc\comp\livehavoc=\fail$: for the same reason as above, $\havoc$
and $\livehavoc$ are also incompatible.
\item $\reqresp\comp\livehavoc=\livehavoc$: this says that $\reqresp$
and $\livehavoc$ are compatible, and in fact that there sequential
composition is equivalent to $\livehavoc$. This is indeed the case,
because, in order to meet the input requirements of $\livehavoc$,
the $\reqresp$ component must ensure that its output is infinitely
often true. The only way for $\reqresp$ to achieve that is to impose
a requirement on its own input, namely, that its own input is infinitely
often true as well. Since the names of input and output variables
do not matter for the property transformer semantics, the result is
identical to the property transformer $\livehavoc$.\end{itemize}
\begin{example}
\textcolor{blue}{\label{ex_checking_refinement_LTL}}Having introduced
guarded systems, we can now give formal semantics to the components
and diagrams introduced in Figures \ref{fig:incomp-1} and \ref{fig:comp-1},
from Example \ref{example_intro}. The semantics of the components
(boxes) in these figures are guarded monotonic property transformers,
defined by LTL formulas. In particular, a component labeled with some
formula $\phi$ corresponds to the guarded transformer $\gsystem{\phi}$.
For instance, component \emph{C} introduced in Figure \ref{fig:comp-1}
corresponds to the guarded transformer $\gsystem{y\leadsto x\mid\always(y=1\Rightarrow\event x=1)}$,
which is equivalent to the transformer $\reqresp$ from Example \ref{ex_guarded}.
(Note that $\reqresp$ uses $x$ as the input and $y$ as the output,
whereas in $C$ it is the other way around. This difference does not
matter, as input and output variables are bound; semantically, the
two systems define identical property transformers.) 

We can also use some of the established results to reason about such
components formally. For example, let us apply the results of Theorem
\ref{lem:assert-demonic} to see how checking refinement of systems
specified in LTL can be reduced to checking satisfiability of quantified
LTL formulas. Consider again Example \ref{example_intro}, and in
particular components $D$ and $E$ from Figure \ref{fig:comp-1}.
Let $\phi_{1}$ be the LTL formula of $D$ and $\phi_{2}$ be the
LTL formula of $E$. Then, checking that $E$ refines $D$ amounts
to checking $\gsystem{\phi_{1}}\sqsubseteq\gsystem{\phi_{2}}$. By
Theorem \ref{lem:assert-demonic}, Part \ref{enu:assert-dem-e}, checking
$\gsystem{\phi_{1}}\sqsubseteq\gsystem{\phi_{2}}$ is equivalent to
checking validity of the formula $\Phi=(\psi_{1}\Rightarrow\psi_{2}\land(\psi_{1}\land\phi_{2}\Rightarrow\phi_{1}))$,
where $\psi_{i}=\inp.\phi_{i}$, for $i=1,2$. The formulas $\psi_{1}$
and $\psi_{2}$ can be obtained from $\phi_{1}$ and $\phi_{2}$ by
existential quantification of the output variables. For example, $\psi_{2}=\inp.\phi_{2}=(\exists x:\always(y=1\Rightarrow x=1)$).
In this specific example, quantifiers can be eliminated in both cases
of $\psi_{1}$ and $\psi_{2}$, and this results in two pure LTL formulas:
$\psi_{1}=\always\event(y=1)$ and $\psi_{2}=\tru$. In general, however,
LTL is not closed under quantifier elimination \cite{Wolper81}. Therefore,
$\Phi$ is generally an LTL formula with quantifiers. Checking validity
of $\Phi$ amounts to checking (un)satisfiability of $\neg\Phi$.
Satisfiability of quantified LTL is decidable, and methods such as
those presented in \cite{SistlaVardiWolper87,KestenPnueli95} can
be used for that purpose.

We can also use some of the established results to reduce checking
compatibility of components to checking satisfiability of formulas
of the appropriate logic, as the following theorem states:\end{example}
\begin{thm}
\label{thm_compatibility}For property transformers $S$ and $T$,
and properties and relations $p$, $p'$, $r$, and $r'$, of appropiate
types, we have:\end{thm}
\begin{enumerate}
\item If $S$ is monotonic, then $S=\fail$ iff $\pfail.S=\top$.
\item If $S$ and $T$ are monotonic, then $S$ and $T$ are incompatible
with respect to $S\comp T$ iff $\pfail.(S\comp T)=\top$.
\item \label{enu_thm_compatibility}$\system pr$ and $\system{p'}{r'}$
are incompatible with respect to $\system pr\comp\system{p'}{r'}$
iff $\neg\exists x:p.x\land(\forall y:r.x.y\Rightarrow p'.y)$.
\item $\gsystem r$ and $\gsystem{r'}$ are incompatible with respect to
$\gsystem r\comp\gsystem{r'}$ iff $\neg\exists x:\inp.r.x\land(\forall y:r.x.y\Rightarrow\inp.r'.y)$.
\end{enumerate}
For instance, Part \ref{enu_thm_compatibility} of Theorem \ref{thm_compatibility}
states that the composition of two relational property transformers
$\gsystem{x\leadsto y\mid P\mid R}\comp\gsystem{y\leadsto z\mid P'\mid R'}$
is valid (i.e., the two are compatible) iff the formula $P\land(\forall y:R\Rightarrow P')$
is satisfiable.

\section{Property transformers based on symbolic transition systems\label{sec_STS}}

So far, we have introduced (relational and guarded) monotonic property
transformers and showed how these can be defined using LTL. As a language,
LTL is often more appropriate for system specification, and less appropriate
for system implementation. For the latter purpose, it is often convenient
to have a language which explicitly refers to state variables and
allows to manipulate them, e.g., by defining the next state based
on the current state and input. In this section we introduce a \emph{symbolic
transition system} notation which allows to do this, and show how
this notation can be given semantics in terms of property transformers. 

For example, suppose we want a counter which accepts as input infinite
sequences of Boolean values, and returns infinite sequences of natural
numbers where every output is the number of true values seen so far
in the input. Moreover, we want this counter to accept inputs where
the number of true values is bounded by a given natural number $n$.
If $\coun.x.i$ is the number of trues in $x{}_{0},\; x_{1},\;\ldots,\; x_{i}$,
then this system can be defined in the following way:
\[
\bcount.n=\assert{x\mid\forall i:\coun.x.i\le n}\comp\demonic xy{\forall i:y_{i}=\coun.x.i}
\]
Although this system is defined globally, when computing $y_{i}$
we only need to know $x_{i}$, and we need to know how many true values
we have seen so far in the input. We can store the number of true
values seen so far in a \emph{state variable} $u$. Then, it would
be natural to define the counter \emph{locally}, that is, define \emph{one
step} of the counter, as follows:
\[
\assert{u_{i}\le n}\comp[u_{i},x_{i}\leadsto u_{i+1},y_{i}\;|\; u_{i+1}=(\ifs{x_{i}}{u_{i}+1}{u_{i}})\land y_{i}=u_{i+1}]
\]
where $i$ is the index of the step, and we can assume that initially
$u_{0}=0$. In the above definition, $u_{i}$ refers to the \emph{current}
state (i.e., the state at current step $i$) and $u_{i+1}$ refers
to the \emph{next} state (i.e., the state at next step $i+1$), while
$x_{i}$ refers to the current input and $y_{i}$ refers to the current
output (both at current step $i$). At every step the assert statement
$\{u_{i}\le n\}$ tests if $u_{i}$ is less or equal to $n$. If this
is false then the system fails because the input requirement that
the number of true values never exceeds $n$ is violated. If $u_{i}\le n$
then we calculate the next state value $u_{i+1}$ and the output value
$y_{i}$. 

Generalizing from this example, a \emph{symbolic transition system}
is a tuple $(init,p,r)$, formed by a predicate $init.u$, a predicate
$p.u.x$, and a relation $r.u.u'.x.y$, where $x$ is the input, $u$
is the current state, $u'$ is the next state, and $y$ is the output.
The predicate $init$ is called the \emph{local initialization predicate}
of the system, the predicate $p$ is called the \emph{local precondition}
of the system, and $r$ is called the \emph{local input-output relation}
of the system. The intuitive interpretation of such a system is that
we start with some initial state $u_{0}\in init$ and we are given
some input sequence $x_{0},\; x_{1},\;\ldots$, and if $p.u_{0}.x_{0}$
is true, then we compute the next state $u_{1}$ and the output $y_{0}$
such that $r.u_{0}.u_{1}.x_{0}.y_{0}$ is true. Next, if $p.x_{1}.u_{1}$is
true, then we compute $u_{2}$ and $y_{1}$ such that $r.u_{1}.u_{2}.x_{1}.y_{1}$
is true, and so on. If at any step $p.u_{i}.x_{i}$ is false, then
the computation fails, and the input $x_{0},\; x_{1},\;\ldots$ is
not accepted. 

Note that the computation defined by the relation $r$ can be nondeterministic
in both next state $u'$ and output $y$. That is, for given values
$x$ and $u$ for the input and current state, there could be multiple
values for the next state $u'$ and output $y$ such that $r.u.u'.x.y$
is true. We must carefully account for this non-determinism when defining
the property transformer based on such a symbolic transition system.
To see the complications that may arise, consider another example:
\begin{eqnarray*}
init.u & = & u\\
p.u.x & = & u\\
r.u.u'.x.y & = & (x=y)
\end{eqnarray*}
In this system, if the current state is true then we choose arbitrarily
a new state $u'$ and we copy the input $x$ into the output $y$.
If the system chooses $u'=\fals$ then in the next step the system
will fail, regardless of the input. This example shows that in a nondeterministic
system, for the same input there could be different choices of internal
states such that in one case the system succeeds while in another
it fails. In the example above the choice of state sequence $(\forall i:u_{i}=\tru)$
results in a successful computation, but all other choices of state
sequences fail. In our definition of property transformers, we accept
an input only if \emph{all} choices of internal states lead to no
failures. 

Formally, we say that an input sequence $x_{0},\; x_{1},\ldots$ is
\emph{illegal} for a symbolic transition system if there is some $k\in\nat$
and some choice $u_{0},\; u_{1},\ldots$ of states and $y_{0},\; y_{1},\;\ldots$
of outputs such that $init.u_{0}$ and $(\forall i<k:r.u_{i}.u_{i+1}.x_{i}.y_{i})$
and $\neg p.u_{k}.x_{k}$. For technical reasons, we need to generalize
$p$ to be a predicate not only on the current state and input, but
also on the next state (the need for this will become clear in the
sequel, see Theorem \ref{thm:localsys-repr} and discussion that follows).
With this generalization, we define the $\illegal$ predicate on symbolic
transition systems and input sequences, as follows:
\[
\illegal.init.p.r.x=(\exists u,y,k:init.u_{0}\land(\forall i<k:r.u_{i}.u_{i+1}.x_{i}.y_{y})\land\neg p.u_{k}.u_{k+1}.x_{k})
\]

We can also formalize a \emph{run} of a symbolic transition system,
using the predicate $\run$. For sequences $x$, $u$, and $y$, the
predicate $\run.r.u.x.y$ is defined by:
\[
\run.r.u.x.y=(\forall i:r.u_{i}.u_{i+1}.x_{i}.y_{i})=\always r.u.\nxt u.x.y
\]
where, recall, $u^{1}$ denotes the sequence $u_{1},u_{2},\cdots$,
i.e., the sequence of states starting from the second state $u_{1}$
instead of the initial state $u_{0}$. If the predicate $\run.r.u.x.y$
is true we say that there is a \emph{run} of the system with the inputs
$x$, the outputs $y$ and the states $u$. We can now define monotonic
property transformers based on symbolic transition systems as follows:
\begin{defn}
\label{def_local_transformer}Consider a symbolic transition system
described by $(init,p,r)$. Such a system defines a monotonic property
transformer called a \emph{local property transformer,} and denoted
$\lsysb{init}pr$, as follows:
\[
\lsysb{init}pr.q.x=\neg\illegal.init.p.r.x\land(\forall u,y:(init.u_{0}\land\run.r.u.x.y)\Rightarrow q.y)
\]
What the above definition states is that an input sequence $x$ is
in the set of input sequences of $\lsysb{init}pr$ that are guaranteed
to establish $q$ iff: (1) $x$ is legal; and (2) for all choices
of state traces $u$ and output traces $y$, if $u_{0}$ satisfies
$init$, and if there is a run of the system with the inputs $x$,
the outputs $y$ and the states $u$, then $y$ must be in $q$.

Before proceeding, let us make a remark on why we use similar, but
different, notation for relational property transformers and for local
property transformers. In the case of relational property transformers
we use notation such as $\gsystem{p\mid r}$. Here, $p$ and $r$
are predicates over \emph{sequences} (traces). For instance, $p$
might be the LTL formula $\always x=1$. In the case of local property
transformers we use notation such as $\lsysb{init}pr$. Here, $init,p$,
and $r$ are local predicates over (input, output, and state) \emph{variables}.
For example, $p$ in this case might be the predicate $u=0\Rightarrow x=1$.
\end{defn}
Note that so far our definition of symbolic transition systems is
essentially semantic, since $init,p,r$ are semantic objects. In practice,
we may use a syntax such as Boolean expressions for these elements.
This is essentially the language used in symbolic model-checking tools
like, say, NuSMV. If $Init,$ $P$, and $R$ are Boolean expressions
possibly containing free the variables $u$ and $u,u',x$ and $u,u',x,y$,
respectively then we define a syntax to describe local property transformers
similar to the syntax we used for relational and guarded property
transformers:
\[
\lsysa{\leadstost xuy}{Init}PR=\lsysb{\lambda u:Init}{\lambda u,u',x:P}{\lambda u,u',x,y:R}
\]

\begin{example}
For instance, we can define the property transformer for the counter
system discussed in the beginning of this section as follows:
\[
\bcount.n=\bsys\leadstost xuy\mid u=0\mid u\le n\mid u'=(\ifs x{u+1}u)\land y=u'\esys
\]
 We can also prove that the non-deterministic example discussed above
is equivalent to the $\fail$ transformer: 
\[
\lsysa{\leadstost xuy}uu{y=x}=\fail
\]
that is, for all input sequences this system fails.\end{example}
\begin{lem}
For a symbolic transition system $(init,p,r)$, the set of input sequences
for which its local property transformer $\lsysb{init}pr$ fails is
equal to the set of its illegal input sequences:

\[
\pfail.\lsysb{init}pr=\illegal.init.p.r
\]

\end{lem}

\subsection{Local property transformers are relational}

The definition of a local property transformer is close to our intuition
of how a system with state should operate, step by step, however,
it is difficult to see immediately from this definition whether local
transformers belong to the class of relational property transformers.
The following theorem shows that this is indeed the case:
\begin{thm}
\label{thm:localsys-repr}For any symbolic transition system $(init,p,r)$,
we have:
\[
\begin{array}{ll}
 & \lsysb{init}pr\\
=\\
 & \demonic x{u,x}{init.u_{0}}\comp\assert{u,x\;|\;(\inp.r\leads p).u.\nxt u.x}\comp\demonic{u,x}y{\always r.u.\nxt u.x.y}\\
=\\
 & \assert{x\;|\;\forall u:init.u_{0}\Rightarrow(\inp.r\leads p).u.\nxt u.x}\comp\demonic xy{\exists u:init.u_{0}\land\always r.u.\nxt u.x.y}
\end{array}
\]

\end{thm}
Theorem \ref{thm:localsys-repr} shows that a local property transformer
$\lsysb{init}pr$ can be expressed as a sequential composition of
assert and update transformers. Since the latter are special cases
of relational transformers, and relational transformers are closed
by sequential composition, this shows that local property transformers
are relational. Moreover, the assert and update transformers used
in the right-hand side of theorem above are constructed by applying
some temporal operators to the \emph{local} precondition $p$ and
the \emph{local} input-output relation $r$. Here, $p$ and $r$ are
\emph{local} in the sense that they refer only to one step, i.e.,
they are predicates on state, input and output variables, and not
on infinite sequences.

Theorem \ref{thm:localsys-repr} also justifies our earlier generalization
of the local precondition to be a function not only on the current
state and the input, but also on the next state. This is so because
the precondition $\inp.r$ depends anyway on the next state $(\inp.r.u.u'.x)$.

We call the precondition $(\inp.r\leads p).u.\nxt u.x$ from the representation
of the local system $\lsysb{init}pr$ the \emph{global precondition}
of $\lsysb{init}pr$. Similarly $\always r.u.\nxt u.x.y$ is the \emph{global
input-output relation} of $\lsysb{init}pr$.

\subsection{Checking that symbolic transition systems refine their specification}

An additional benefit of Theorem \ref{thm:localsys-repr} is that
it makes checking refinement of local systems against their specification
(or against another system) easier. For example, suppose that we want
to prove a refinement like
\[
\system pr\sqsubseteq\lsysb{init}{p'}{r'}
\]
If we use the original definition of local system (Definition \ref{def_local_transformer}),
then we need to expand the definition of $\lsysb{init}{p'}{r'}$ and
reason about individual values of traces ($x_{i}$, $y_{i}$, $u_{i}$).
This reasoning is at a lower level than for example the reasoning
about the refinement
\[
\assert p\comp\assume r\sqsubseteq\assert{p''}\comp\assume{r''}
\]
which, by Theorem \ref{lem:assert-demonic}, is equivalent to 
\[
(\forall x:p.x\Rightarrow p''.x)\land(\forall x,y:p.x\land r''.x.y\Rightarrow r.x.y)
\]
In this property $x$, $y$ may also stand for traces, but this formula
does not contain references to specific values ($x_{i}$ or $y_{i}$)
of these traces. Therefore, checking validity of this formula can
be often reduced to simpler problems, e.g., satisfiability of LTL
formulas, as explained in Example \ref{ex_checking_refinement_LTL}.

We can exploit Theorem \ref{thm:localsys-repr} to obtain an analogous
result for local transformers:
\begin{thm}
For $init,\; p,\; p',\; r,\; r'$ of appropriate types we have:
\[
\begin{array}{ll}
 & \system{x\leadsto y\;|\; p}r\sqsubseteq\lsysb{init}{p'}{r'}\\
\Leftrightarrow\\
 & (\forall u,x:init.u_{0}\land p.x\Rightarrow(\inp.r'\leads p').u.\nxt u.x)\land(\forall u,x,y:init.u_{0}\land p.x\land\always r'.u.\nxt u.x.y\Rightarrow r.x.y)
\end{array}
\]

\end{thm}

\subsection{Sequential composition of local transformers}

Theorems \ref{lem:assert-demonic} and \ref{thm:localsys-repr} allow
us to calculate the sequential composition of two local systems, as
follows:
\begin{thm}
\label{thm:local-sys-comp}~
\[
\begin{array}{ll}
 & \lsysb{init}pr\comp\lsysb{init'}{p'}{r'}\\
=\\
 & \assert{x\mid\forall u,v:init.u_{0}\land init'.v_{0}\Rightarrow(\inp.r\leads p).u.\nxt u.x\land(\forall y:\always r.u.\nxt u.x.y\Rightarrow(\inp.r'\leads p').v.\nxt v.y)}\comp\\
 & \qquad\demonic xz{\exists u,v:init.u_{0}\land init'.v_{0}\land\always(r\relstcomp r').(u,v).(\nxt u,\nxt v).x.z}
\end{array}
\]
where $(r\relstcomp r').(u,v).(u',v')=(r.u.u'\circ r'.v.v')$
\end{thm}
Ideally the composition of two local systems $S=\lsysb{init}pr$ and
$S'=\lsysb{init'}{p'}{r'}$ would be a local system corresponding
to the composition of the local transitions of $S$ and $S'$. Unfortunately
this is not the case. In the rest of this subsection we explain why
this is the case. This will motivate the definition of a restricted
class of local transformers, called \emph{guarded} local transformers,
which are analogous to guarded property transformers, and enjoy good
closure properties.

We begin by defining the \emph{local transition} of a local system
to be the predicate transformer: 
\[
\ltran.p.r=\assert{x,u\;|\; p.u.x}\comp\demonic{x,u}{y,u'}{r.u.u'.x.y}
\]
Note that $\ltran.p.r$ is a \emph{predicate} transformer, not a property
transformer. Also note that here it suffices to consider $p$ as a
predicate on the current state and input only. The execution of $\ltran.p.r$
starts from the input value $x$ and the state $u$ and if $p.u.x$
is true, then it computes the output value $y$ and the new state
$u'$ such that $r.u.u'.x.y$ is true. The local transitions of $S$
and $S'$ have the local states $u$ and $v$, respectively, and their
compostion will have the local state pairs $(u,v)$. In order to be
able to compose the local transitions of $S$ and $S'$, we add the
state $v$ to the local transition of $S$ and the state $u$ to local
transition of $S'$: 

\begin{eqnarray*}
\ltran\mbox{*}.p.r & = & \assert{x,u,v\;|\; p.u.x}\comp\demonic{x,u,v}{y,u',v}{r.u.u'.x.y}\\
\ltran\mbox{*}.p'.r' & = & \assert{y,u,v\;|\; p'.v.y}\comp\demonic{y,u,v}{z,u,v'}{r'.v.v'.y.z}
\end{eqnarray*}
We show what would be the local system for the composition of the
local transitions of $S$ and $S'$. We have
\[
\begin{array}{ll}
 & \ltran\mbox{*}.p.r\comp\ltran\mbox{*}.p'.r'\\
= & \mbox{\{Theorem \ref{lem:assert-demonic}\}}\\
 & \assert{x,u,v\;|\; p.u.x\land(\forall u',y:r.u.u'.x.y\Rightarrow p'.v.y)}\comp\demonic{x,u,v}{z,u,v'}{(r.u.u')\circ(r'.v.v').x.z}
\end{array}
\]
 The sequential composition of the two systems has as state pairs
$(u,v)$ and the initialization predicate, the local precondition,
and the local relation of the composition should be given by 
\begin{eqnarray*}
init''.(u,v) & = & init.u\land init'.v\\
p''.(u,v).x & = & p.u.x\land(\forall u',y:r.u.u'.x.y\Rightarrow p'.v.y)\\
r'' & = & r\relstcomp r'
\end{eqnarray*}
The local system of the composition of the local transitions of $S$
and $S'$ is
\[
\begin{array}{ll}
 & \lsysb{init''}{p''}{r''}\\
=\\
 & \assert{x\;|\;\forall u,v:init''.(u_{0},v_{0})\Rightarrow(\inp.r''\leads p'').(u,v).(\nxt u,\nxt v).x}\comp\\
 & \qquad\demonic xz{\exists u,v:init''.(u_{0},v_{0})\land\always r''.(u,v).(\nxt u,\nxt v).x.z}
\end{array}
\]
Now, one might expect the equality $S\comp S'=\lsysb{init''}{p''}{r''}$.
Unfortunately this does not generally hold for arbitrary local systems
$S$ and $S'$. We do have, by definition
\[
init''.(u,v)=init.u\land init'.v\mbox{ and }r''=r\relstcomp r'
\]
but there exist $p,\; r,\; p',\; r',\; u,\; v,$ and $x$ such that
\begin{equation}
\begin{array}{ll}
 & (\inp.r''\leads p'').(u,v).(\nxt u,\nxt v).x\\
\not=\\
 & (\inp.r\leads p).u.\nxt u.x\land(\forall y:\always r.u.\nxt u.x.y\Rightarrow(\inp.r'\leads p').v.\nxt v.y)
\end{array}\label{eq:prec-local-comp}
\end{equation}
For example, if we take
\[
\begin{array}{lll}
init.u & = & (u=0)\\
p.u.x & = & \tru\\
r.u.u'.x.y & = & (u=0\land u'=1)\\
init'.v & = & \tru\\
p'.v.y & = & \fals\\
r'.v.v'.y.z & = & \tru
\end{array}
\]
then (\ref{eq:prec-local-comp}) becomes true. 

What happens in this case is that $S=\magic$ so $S\comp S'=\magic$,
whereas $\lsysb{init''}{p''}{r''}=\fail$, and therefore clearly $S\comp S'\not=\lsysb{init''}{p''}{r''}$.
Intuitively, when executing the system $S\comp S'$, the precondition
$p'$ of $S'$ is tested after a complete execution of $S$, however
in our example above, the execution of $S$ proceeds normally with
the first step when started in the state $u=0$, but then next step
is miraculous because $r.1.u'.x.y$ is false. Therefore the assertion
of $S'$ containing $p'$ is not reached. On the other hand the execution
of $\lsysb{init''}{p''}{r''}$ starting from the same initial state
$u=0$ proceeds normally with the first step of $S$ ($\assert{x,u\;|\; p.u.x}\comp\demonic{x,u}{y,u'}{r.u.u'.x.y}$),
and then tests $p'$, and it fails because $p'$ is false.

\subsection{Guarded local systems}

As we have seen from the previous section, the composition of two
local tranformers is not necessarily a local transformer. This is
because of the possible miraculous behavior of such systems. In this
section we restrict the local precondition of a local system such
that we do not have miraculous behavior anymore. We achieve this by
considering systems where the local precondition $p$ is $\inp.r$.
This is similar to what we have done in order to restrict general
relational transformers to guarded systems, in Section \ref{sub:Guarded-systems}.
\begin{defn}
The\emph{ guarded local system} of $init$ and $r$ is denoted by
$\lsysc{init}r$ and it is given by
\[
\lsysc{init}r=\lsysb{init}{\inp.r}r
\]
and the \emph{local precondition} of a local guarded reactive systems
with state is $\inp.r$.\end{defn}
\begin{thm}
\label{thm:glocalsys-repr}For $init$ and $r$ as in the definition
of a local guarded system we have:
\[
\begin{array}{l}
\lsysc{init}r=\demonic x{u,x}{init.u_{0}}\comp\gsystem{u,x\leadsto y\;|\;\always r.u.\nxt u.x.y}\end{array}
\]

\end{thm}
The next theorem shows that the sequential composition of two local
guarded systems is also a local guarded system.
\begin{thm}
For $init$, $init'$, $r$, and $r'$ we have
\[
\begin{array}{l}
\lsysc{init}r\comp\lsysc{init'}{r'}=\lsysc{init''}{rel\_comp.r.r'}\end{array}
\]
where
\[
init''.(u,v)=init.u\land init'.v
\]
and
\[
\begin{array}{ll}
 & rel\_comp.r.r'.(u,v).(u',v').x.z\\
=\\
 & (\inp.r.u.u'.x\land(\forall y:r.u.u'.x.y\Rightarrow\inp.r'.v.v'.y)\land((r.u.u')\circ(r'.v.v')).x.z)
\end{array}
\]

\end{thm}

\subsection{Stateless systems}

We define \emph{stateless systems} as a special case of local systems,
where the state $u$ ranges over a singleton set $\{\bullet\}$ and
where $init.u=\tru$. In this case we have
\begin{thm}
\label{thm:less-localsys-repr}For $p$, and $r$ as in the definition
of a stateless system, we have:
\begin{eqnarray*}
\lsysb{init}pr & = & \system{x\leadsto y\;|\;(\inp.r\leads p).x}{(\always r).x.y}\\
 & = & \system{\inp.r\leads p}{\always r}
\end{eqnarray*}

\end{thm}
Based on this theorem we use the notation $\system{\inp.r\leads p}{\always r}$
for a stateless system. 

The next theorem gives a procedure to calculate the sequential composition
of two stateless systems.
\begin{thm}
\label{thm:less-local-sys-comp}~
\[
\begin{array}{ll}
 & \system{\inp.r\leads p}{\always r}\comp\system{\inp.r'\leads p'}{\always r'}\\
=\\
 & \assert{x\;|\;(\inp.r\leads p).x\land(\forall y:\always r.x.y\Rightarrow(\inp.r'\leads p').y)}\comp\assume{\always(r\circ r')}
\end{array}
\]

\end{thm}
As in the case of general local systems, the composition of two stateless
systems is not always a stateless system. This motivates us to introduce
guarded stateless systems, similarly to guarded local systems.

\subsection{Guarded stateless systems}
\begin{defn}
A \emph{guarded stateless system} is a stateless system where $p=\inp.r$.\end{defn}
\begin{thm}
\label{thm:glocalsys-repr-1}For any $r$: $\system{\inp.r\leads\inp.r}{\always r}=\gsystem{\always r}$
.
\end{thm}
This is because we have 
\[
\system{\inp.r\leads\inp.r}{\always r}=\system{\always\inp.r}{\always r}=\system{\inp.(\always r)}{\always r}=\gsystem{\always r}
\]
We use the notation $\gsystem{\always r}$ for a guarded stateless
system.

Sequential composition of guarded stateless systems is also a guarded
stateless system:
\begin{thm}
For $r$, and $r'$ we have
\[
\gsystem{\always r}\comp\gsystem{\always r'}=\gsystem{\always(rel\_comp.r.r')}
\]
where $rel\_comp$ is as defined before, but without the state parameters
$u,\; u',\; v,$ and $v'$.
\end{thm}

\subsubsection{Local properties}

Next we introduce some special properties and we show that stateless
guarded local systems behave consistently with respect to these properties.
\begin{defn}
For a property $q$, the $i$-th \emph{projection} of $q$ is a predicate
on states given by
\[
\proj.q.i.s=(\exists\sigma:q.\sigma\land\sigma_{i}=s)
\]
and a property $q$ is a \emph{piecewise local property} if it satisfies
the condition
\[
(\forall\sigma:(\forall i:\proj.q.i.\sigma_{i})\Rightarrow q.\sigma)
\]

\end{defn}
Equivalently, a property $q$ is piecewise local if there exist some
predicates $p_{0}$, $p_{1}$, $\ldots$ such that 
\[
(\forall\sigma:\sigma\in q\Leftrightarrow(\forall i:p_{i}.\sigma_{i}))
\]
There are properties which are not local. For example the liveness
property $q=(\always(\event x))$ is not local because $\sigma=(\lambda i:\fals)$
satisfies the condition $(\forall i:\proj.q.i.(\sigma.i))$, but $\sigma\not\in q$.
\begin{lem}
If $r$ is a state relation, then\end{lem}
\begin{enumerate}
\item $\gsystem{\always r}.q.x\Rightarrow(\forall i:\gsystem r.(\proj.q.i).x_{i})$
\item If $q$ is piecewise local then $(\forall i:\gsystem r.(\proj.q.i).x_{i})\Rightarrow\gsystem{\always r}.q.x$.
\end{enumerate}
This lemma asserts that for the piecewise local property $q$ and
input sequence $x$, all possible outputs of $\gsystem{\always r}$
starting from $x$ are in $q$ if and only if for all steps $i$ all
possible outputs of $\gsystem r$ from $x_{i}$ are in $\proj.q.i$.
So the global execution of $\gsystem{\always r}$ is equivalent to
the execution of $\gsystem r$ on all steps. 
\begin{example}
If we have a local system it does not necessarily mean that we cannot
study its behavior with respect to non piecewise local properties.
For example let us consider a stateless local guarded system that
at each step computes $y=x$ or $y=x+1$, assuming that $x>0$: 
\[
S=\gsystem{x\leadsto y\mid\always(x>0\land(y=x\lor y=x+1))}
\]
If we want to see under what conditions on the input $x$ the output
of $S$ satisfies the property $q=\always\event y<10$, then we should
calculate $S.q$:
\[
S.q=\always(x>0\land\event x<9)
\]
If we want to see under what conditions on the input the output of
$S$ satisfies $q'=\always\event y=10$, then we should calculate
$S.q'$. In this case we have $S.q'=\bot$. This is so because of
the demonic choice $y=x$ or $y=x+1$. For all values of $x$ it is
always possible to choose $y\not=10$.
\end{example}

\section{Application: extending relational interfaces with liveness}

To illustrate the power of our framework, we show how it can handle
as a special case the extension of the relational interface theory
presented in \cite{tripakis:2011} to infinite behaviors and liveness.
We note that the theory proposed in \cite{tripakis:2011} allows to
describe only safety properties, in fact, finite and prefix-closed
behaviors. Extending to infinite behaviors and liveness properties
is mentioned as an open problem in \cite{tripakis:2011}. 

A number of examples showcasing this extension have already been provided
in the introduction. Here we provide an additional example. Consider
the following symbolic transition system:
\[
\begin{array}{rll}
init.u & = & (u=0)\\
p.u.u'.x & = & (-1\le u\le3)\\
r.u.u'.x.y & = & ((x\land u'=u+1)\lor(\neg x\land u'=u-1)\lor u'=0)\land y=(u'=0))
\end{array}
\]

This symbolic transition system has a Boolean input $x$ and a Boolean
output $y$. If the input is true then state counter $u$ is incremented.
If the input is false then $u$ is decremented. Regardless of the
input, the system may also choose nondeterministically to reset the
counter to zero. The output of the system is true whenever the counter
reaches zero. The system also restricts the value of the state to
be between $-1$ and $3$. If the state goes out of this range the
system will fail. The system is supposed to start from state $u=0$.
The local system for this relation is
\begin{equation}
\begin{array}{ll}
 & \lsysb{init}pr\\
=\\
 & \assert{x\mid\forall u:init.u_{0}\Rightarrow(\inp.r\leads p).u.\nxt u.x}\comp\assume{x\leadsto y\mid\exists u:init.u_{0}\land\always r.u.\nxt u.x.y}
\end{array}\label{eq:ex-no-live}
\end{equation}
However we are interested in a system which is also capable of ensuring
the liveness property that $y$ is true infinitely often. We achieve
this by adding the constraint $\always\event y$ to the input-output
relation of (\ref{eq:ex-no-live}). So the full example is 
\begin{equation}
\begin{array}{ll}
 & \mathsf{EXAMPLE}\\
=\\
 & \assert{x\mid\forall u:init.u_{0}\Rightarrow(\inp.r\leads p).u.\nxt u.x}\comp\assume{x\leadsto y\mid\exists u:init.u_{0}\land\always r.u.\nxt u.x.y\land\always\event y}
\end{array}
\end{equation}
In this example the state condition $-1\le u\le3$ is a safety property,
and we designed the example such that this property is enforced on
the input. That is, some input trace is accepted by this system only
if this property is not violated. For example the input sequence $x_{0}=\tru$,
$x_{1}=\fals$, $\ldots$ maintains this property. On the other hand
the property $\always\event y$ is a liveness property which is guaranteed
by the system, regardless of the input.  If we need we can move this
property to the precondition (adapted to the state variable) and then
the system will fail if the input is such that this property is false.
We can prove that our example system establishes the liveness property
$\always\event y$ for all inputs that do not fail, i.e., for all
input traces which satisfy
\[
prec\_g.x=(\forall u:init.u_{0}\Rightarrow(\inp.r\leads p).u.\nxt u.x)
\]
We have
\begin{equation}
\forall x:\mathsf{EXAMPLE.}(\{y\mid\always\event y\}).x=prec\_g.x\label{eq:liveness}
\end{equation}

We can now use $\mathsf{EXAMPLE}$ as specification and we can, for
instance, refine it to the system which always assigns $\tru$ to
the output variable:
\[
\mathsf{EXAMPLE}\sqsubseteq\assume{x\leadsto y\;|\;\always y}.
\]
 We can also assume that the input satisfies some additional property.
For instance, we can assume that $x$ is alternating between $\tru$
and $\fals$:
\[
\assert{x\mid\always(x\Leftrightarrow\neg\nex x)}\comp\mathsf{EXAMPLE}
\]
Then we can show that this new system is refined by the original symbolic
transition system:
\[
\begin{array}{ll}
 & \assert{x\mid\always(x\Leftrightarrow\neg\nex x)}\comp\mathsf{EXAMPLE}\\
\sqsubseteq\\
 & \assert{x\mid\always(x\Leftrightarrow\neg\nex x)}\comp\lsysb{init}pr\\
\sqsubseteq\\
 & \lsysb{init}pr
\end{array}
\]
because the additional property used as precondition ensures the liveness
property.

Using this formalism we can construct liveness specifications as the
example system, and we can refine them in appropriate contexts to
systems which do not have any liveness property, but they preserve
the liveness property of the input. 

From (\ref{eq:liveness}) we also obtain 
\[
\mathsf{EXAMPLE}=\mathsf{EXAMPLE}\comp\assert{y\mid\always\event y}
\]
We can use this property when constructing another system that uses
the output from $\mathsf{EXAMPLE}$ as input. Then we know that this
input satisfies the liveness property $\always\event y$ and we can
design this second system accordingly.

\section{Conclusions}

In this paper we introduced a monotonic property transformer semantics
for reactive systems. The semantics supports refinement, compostion,
compatibility, demonic choice, unbounded nondeterminism, and other
interesting system properties. The semantics also supports angelic
choice: we have not specifically exploited this feature here and we
leave it for future work. The semantics can be used to specify and
reason about both safety and liveness properties. Our framework allows
to describe systems using a variety of formalisms, from higher order
logic, to temporal logic, to symbolic transition systems. The framework
is compositional, in particular if we restrict ourselves to the most
realistic case of guarded systems, which cannot behave miraculously,
and enjoy good closure properties. Our work generalizes previous work
on relational interfaces to systems with infinite behavior and liveness
properties. Future work includes studying more operators (e.g., angelic
choice), and extending the framework to continuous-time and hybrid
systems.

\bibliographystyle{plain}

\begin{thebibliography}{10}

\bibitem{AlurHenzingerFMSD99}
Rajeev Alur and Thomas~A. Henzinger.
\newblock Reactive modules.
\newblock {\em Formal Methods in System Design}, 15:7--48, 1999.

\bibitem{back-1978}
Ralph-Johan Back.
\newblock {\em On the correctness of refinement in program development}.
\newblock {PhD} thesis, Department of Computer Science, University of Helsinki,
  1978.

\bibitem{back:1990}
Ralph-Johan Back.
\newblock {Refinement calculus, part II: Parallel and reactive programs}.
\newblock In {\em Stepwise Refinement of Distributed Systems Models,
  Formalisms, Correctness}, pages 67--93. Springer, 1990.

\bibitem{back-wright-98}
Ralph-Johan Back and Joakim von Wright.
\newblock {\em Refinement Calculus. A systematic Introduction}.
\newblock Springer, 1998.

\bibitem{back:wright:1994}
Ralph-Johan Back and Joakim Wright.
\newblock Trace refinement of action systems.
\newblock In Bengt Jonsson and Joachim Parrow, editors, {\em CONCUR '94:
  Concurrency Theory}, volume 836 of {\em Lecture Notes in Computer Science},
  pages 367--384. Springer Berlin Heidelberg, 1994.

\bibitem{back:xu:1998}
Ralph-Johan Back and Qiwen Xu.
\newblock Refinement of fair action systems.
\newblock {\em Acta Informatica}, 35(2):131--165, 1998.

\bibitem{BroyStolen01}
Manfred Broy and Ketil St{\o}len.
\newblock {\em Specification and development of interactive systems: focus on
  streams, interfaces, and refinement}.
\newblock Springer, 2001.

\bibitem{AlfaroHenzingerFSE01}
Luca de~Alfaro and Thomas~A. Henzinger.
\newblock Interface automata.
\newblock In {\em Foundations of Software Engineering (FSE)}. ACM Press, 2001.

\bibitem{dill:1989}
David~L. Dill.
\newblock {\em Trace Theory for Automatic Hierarchical Verification of
  Speed-independent Circuits}.
\newblock MIT Press, Cambridge, MA, USA, 1989.

\bibitem{Harel:1989:DRS:101969.101990}
David Harel and Amir Pnueli.
\newblock On the development of reactive systems.
\newblock In Krzysztof~R. Apt, editor, {\em Logics and Models of Concurrent
  Systems}, pages 477--498. 1985.

\bibitem{KestenPnueli95}
Yonit Kesten and Amir Pnueli.
\newblock {A complete proof system for QPTL}.
\newblock In {\em LICS}, June 1995.

\bibitem{LynchTuttle89}
Nancy~A. Lynch and Mark~R. Tuttle.
\newblock An introduction to input/output automata.
\newblock {\em CWI Quarterly}, 2:219--246, 1989.

\bibitem{nipkow-paulson-wenzel-02}
Tobias Nipkow, Lawrence~C. Paulson, and Markus Wenzel.
\newblock {\em {Isabelle/HOL} --- A Proof Assistant for Higher-Order Logic},
  volume 2283 of {\em LNCS}.
\newblock Springer, 2002.

\bibitem{pnueli:1977}
Amir Pnueli.
\newblock The temporal logic of programs.
\newblock In {\em Foundations of Computer Science, 1977., 18th Annual Symposium
  on}, pages 46--57, Oct 1977.

\bibitem{preoteasa:2014afp}
Viorel Preoteasa.
\newblock Formalization of refinement calculus for reactive systems.
\newblock {\em Archive of Formal Proofs}, June 2014.
\newblock \url{http://afp.sf.net/entries/RefinementReactive.shtml}, Formal
  proof development. {\em Review pending}.

\bibitem{preoteasa:2013}
Viorel Preoteasa.
\newblock Refinement algebra with dual operator.
\newblock {\em Science of Computer Programming}, 92, Part B(0):179 -- 210,
  2014.
\newblock Selected papers from the Brazilian Symposium on Formal Methods (SBMF
  2011).

\bibitem{SistlaVardiWolper87}
Prasad Sistla, Moshe~Y. Vardi, and Pierre Wolper.
\newblock The complementation problem for {B}\"uchi automata with applications
  to temporal logic.
\newblock {\em Theoretical Computer Science}, 49:217--237, 1987.

\bibitem{tripakis:2011}
Stavros Tripakis, Ben Lickly, Thomas~A. Henzinger, and Edward~A. Lee.
\newblock A theory of synchronous relational interfaces.
\newblock {\em ACM Trans. Program. Lang. Syst.}, 33(4):14:1--14:41, July 2011.

\bibitem{Wolper81}
Pierre Wolper.
\newblock Temporal logic can be more expressive.
\newblock In {\em Foundations of Computer Science}, pages 340--348, 1981.

\end{thebibliography}

\end{document}